\documentclass[11pt]{article}
\usepackage{latexsym}
\usepackage{amssymb}

\usepackage{amsmath,amsfonts,amssymb,caption,graphicx,natbib}
\input{epsf}

\newcommand{\Ahat}{\hat{A}}

\newcommand{\Dmax}{\mbox{$D_{\rm max}$}}

\newcommand{\Dbar}{\mbox{$\overline{D}$}}

\def\be{\begin{eqnarray}}
\def\ee{\end{eqnarray}}
\def\ben{\begin{eqnarray*}}
\def\een{\end{eqnarray*}}

\def\flabel#1{\label{f:#1}}
\def\elabel#1{\label{e:#1}}

\def\sq{$\Box$}

\def\qed{\ifmmode\sq\else{\unskip\nobreak\hfil
\penalty50\hskip1em\null\nobreak\hfil\sq
\parfillskip=0pt\finalhyphendemerits=0\endgraf}\fi\par\medbreak}

\newsavebox{\junk}
\savebox{\junk}[1.6mm]{\hbox{$|\!|\!|$}}

\def\limsup{\mathop{\rm lim\ sup}}

\def\til={{\widetilde =}}

 \def\eq#1/{(\ref{#1})}

\def\eq#1/{(\ref{e:#1})}

\newcommand{\beqn}[1]{\notes{#1}%
\begin{eqnarray} \elabel{#1}}

\newcommand{\eeqn}{\end{eqnarray} }

\newcommand{\beq}[1]{\notes{#1}%
\begin{equation}\elabel{#1}}

\newcommand{\eeq}{\end{equation}} 

\def\bdes{\begin{description}}
\def\edes{\end{description}}

\def\notes#1{}

\input{epsf}

\title{Lossy Compression in Near-Linear Time\\
via Efficient Random Codebooks and Databases}

\author
{
    Christos Gioran
    \thanks{Department of Informatics,
        Athens University of Economics and Business,
        Patission 76, Athens 10434, Greece.
                Email: {\tt himicos@gmail.com}.
        }
\and
        Ioannis Kontoyiannis,
	{\sl Senior Member, IEEE}
    \thanks{Department of Informatics,
        Athens University of Economics and Business,
        Patission 76, Athens 10434, Greece.
                Email: {\tt yiannis@aueb.gr}.
		Web: {\tt http://pages.cs.aueb.gr/users/yiannisk/}.
        }
}

\date{\today}

\begin{document}
\bibliographystyle{plain}
\maketitle

\begin{abstract}
The compression-complexity trade-off of lossy 
compression algorithms that are based on a random codebook or 
a random database is examined.  Motivated, in part, by recent 
results of Gupta-Verd\'{u}-Weissman (GVW) and their underlying 
connections with the pattern-matching scheme of Kontoyiannis' 
lossy Lempel-Ziv algorithm, we introduce a non-universal 
version of the lossy Lempel-Ziv method 
(termed LLZ). The optimality of LLZ for memoryless sources
is established, and its performance is compared to that 
of the GVW divide-and-conquer approach. Experimental results 
indicate that the GVW approach often yields better compression 
than LLZ, but at the price of much higher memory requirements. 
To combine the advantages of both, we introduce a hybrid 
algorithm (HYB) that utilizes both the divide-and-conquer 
idea of GVW and the single-database structure of LLZ. 
It is proved that HYB shares with GVW the exact same rate-distortion 
performance and implementation complexity, while,
like LLZ, requiring less memory, by a factor which may
become unbounded, depending on the choice or the
relevant design parameters.
Experimental results are also presented, illustrating the 
performance of all three methods on data generated by simple 
discrete memoryless sources. In particular, the HYB algorithm is shown 
to outperform existing schemes for the compression of some
simple discrete sources with respect to the Hamming distortion 
criterion.
\end{abstract}

\noindent
{\small
{\bf Keywords --- } 
Lossy data compression, rate-distortion theory,
pattern matching, Lempel-Ziv, random codebook,
fixed database}

\footnotetext{Preliminary versions of the present
results were presented at the IEEE Information
Theory Workshop, Volos, Greece, June 2009.}

\section{Introduction}
\label{s:intro}
%%%%%%%%%%%%%%%%%%%%%%%%%%%%%%%%%%%%%%%%%%%%%%%%%%%%%%%%%%%%%%%%%%

One of the last major outstanding classical problems 
of information theory is the development of 
general-purpose, practical, efficiently implementable
lossy compression algorithms. The corresponding
problem for lossless data compression was essentially
settled in the late 1970s by the advance of
the Lempel-Ziv (LZ) family of algorithms 
\cite{ziv-lempel:1}\cite{ziv:2}\cite{ziv-lempel:2}
and arithmetic coding
\cite{rissanen:76}\cite{pasco:76}\cite{langdon:84};
see also the texts 
\cite{hankerson:book}\cite{bell:cleary:witten}.
Similarly, from the early- to mid-1990s on, 
efficient channel coding strategies emerged
that perform close to capacity, primarily using 
sparse graph codes, turbo codes, and local
message-passing decoding algorithms;
see, e.g., 
\cite{wiberg:95}\cite{mackay:95}\cite{sipser:96}\cite{mao:00},
the texts \cite{frey:book}\cite{mackay:book}\cite{urbanke:book},
and the references therein.

For lossy data compression, although there is a rich 
and varied literature on both theoretical results 
and practical compression schemes, 
near-optimal, efficiently implementable algorithms
are yet to be discovered.
From rate-distortion theory \cite{berger:book}\cite{shannon:59}
we know that it is possible to achieve a sometimes dramatic 
improvement in compression performance by allowing for
a certain amount of distortion in the reconstructed data. 
But the majority of existing algorithms are either
compression-suboptimal or they involve 
exhaustive searches of 
exponential complexity at the encoder,
making them unsuitable for realistic 
practical implementation.

Until the late 1990s, most of the research
effort was devoted to addressing the issue
of universality, see \cite{kieffer:93} and the 
references therein, as well as 
\cite{ziv:II}\cite{neuhoff-gray-davisson}\cite{ziv:3}\cite{ornstein-shields}%
\cite{muramatsu-kanaya}\cite{zhang-wei:1}\cite{zhang-yang:96}%
\cite{yang-kieffer:2};
algorithms emphasizing more practical aspects
have been proposed in \cite{yang:zhang:berger}.
In addition to many application-specific families
of compression standards (e.g., JPEG for images
and MPEG for video), there is a general theory
of algorithm design based on vector quantization;
see
\cite{gray-gersho:book}\cite{linder-lugosi-zeger}%
\cite{chou-effros-gray}\cite{gray-neuhoff}
and the references therein.
Yet another line of research, closer in spirit
to the present work, is on lossy extensions of
the celebrated Lempel-Ziv schemes, based on 
approximate pattern matching; see
\cite{morita-kobayashi:1}\cite{steinberg-gutman}%
\cite{zamir-rose:01}\cite{luczak-szpankowski}%
\cite{arnaud-szpankowski}\cite{yang-kieffer:1}%
\cite{a-g-spa:99}\cite{dembo-kontoyiannis}%
\cite{kontoyiannis-lossy1-1}\cite{a-spa-g:02}.

More recently, there has been renewed interest in
the compression-complexity trade-off, and in the 
development of low-complexity compressors that 
give near-optimal performance, at least for simple
sources with known statistics.
The lossy LZ algorithm of 
\cite{kontoyiannis-lossy1-1} is 
rate-distortion optimal and of polynomial
complexity, although, in part due the penalty
paid for universality,
its convergence is slow.
For the uniform Bernoulli source, 
\cite{matsunaga-yamamoto:03}\cite{miyake:06}\cite{martinian-wainwright:06}
present codes based on sparse graphs, and, although their performance 
is promising, like earlier approaches they rely on exponential searches at the
encoder. In related work, \cite{maneva-wainwright:05}\cite{ciliberti-et-al:06}
present sparse-graph compression schemes with much more attractive
complexity characteristics, but suboptimal compression performance.
Rissanen and Tabus \cite{rissanen-tabus} describe a different method which, 
unlike most of the earlier approaches,  is not based on a random 
(or otherwise exponentially large) codebook. It has linear complexity 
in the encoder and 
decoder and, although it appears to be rate-distortion suboptimal, it is 
an effective practical scheme for Bernoulli sources.
Sparse-graph codes that are compression-optimal and of subexponential
complexity are constructed in \cite{gupta-verdu:pre}. A simulation-based
iterative algorithm is presented in \cite{jalali-weissman:isit08} 
and it is
shown to be compression-optimal, although its complexity is hard to 
evaluate precisely as it depends on the convergence of a Markov chain
Monte Carlo sampler. The more recent work \cite{jalali-et-al:09} on the
lossy compression of discrete Markov sources also contains
promising results;
it is based on the combination of a Viterbi-like optimization
algorithm at the encoder followed by universal lossless 
compression. 

The present work is partly motivated by the 
results reported in \cite{gupta-et-al:pre:09}
by Gupta-Verd\'{u}-Weissman (GVW). 
Their compression schemes 
are based on the ``divide-and-conquer'' approach,
namely the idea that instead of encoding a long 
message $x_1^n=(x_1,x_2,\ldots,x_n)$ using a classical 
random codebook of blocklength $n$, it is preferable
to break up $x_1^n$ into shorter sub-blocks of 
shorter length $\ell$, say, and encode the sub-blocks
separately. The main results in \cite{gupta-et-al:pre:09}
state that, with an appropriately chosen
sub-block length $\ell$, it is possible to achieve
asymptotically optimal rate-distortion performance
with near-linear implementation complexity
(in a sense made precise in Section~\ref{s:hyb} below).

Our starting point is the observation that there 
is a closely related, in a sense dual, point of view. 
On a conceptual as well as mathematical level,
the divide-and-conquer approach is very closely related
to a pattern-matching scheme with a restricted database.
In the divide-and-conquer setting, given a target 
distortion level $D$ and an $\ell\geq 1$,
each sub-block of length $\ell$ in the original message 
$x_1^n$ is encoded using a random codebook consisting 
of $\approx 2^{\ell R(D)}$ codewords, where
$R(D)$ is the rate-distortion function of the source
being compressed (see the following section for 
more details and rigorous definitions).
To encode each sub-block, the encoder searches all
$2^{\ell R(D)}$ entries of the codebook,
in order to find the one which has the smallest 
distortion with respect to that sub-block.

Now suppose that, instead of a random codebook,
the encoder and decoder share
a random database with length 
$M\approx 2^{\ell R(D)}$, generated from the same 
distribution as the Shannon-optimal codebook.
As in \cite{kontoyiannis-lossy1-1}, the encoder 
searches for the longest prefix $x_1^L=(x_1,x_2,\ldots,x_L)$
of the message $x_1^n$ that matches somewhere in the
database with distortion $D$ or less. 
Then the prefix $x_1^L$ is described to the decoder
by describing the position and length of the match
in the database, and the same process is repeated
inductively starting at $x_{L+1}$.
Although the match-length $L$ is random, we know 
\cite{dembo-kontoyiannis}\cite{kontoyiannis-lossy1-1}
that, asymptotically, it behaves like,
$$L\approx \frac{\log M}{R(D)} \approx \ell,
\;\;\;\;
\mbox{with high probability}.$$
Therefore, because the length $M$
of the database was chosen to be $\approx 2^{\ell R(D)}$, 
in effect both schemes will individually 
encode sub-blocks of approximately
the same length $\ell$,
and will also have comparable implementation
complexity at the encoder.\footnote{It is 
	well-known that the main difficulty
	in designing effective lossy compressors
	is in the implementation complexity
	of the {\em encoder}. Therefore,
	in all subsequent results dealing with
	complexity issues we focus on the case
	of the encoder. Moreover, it is easy to
	see that the decoding complexity of all
	the schemes considered here is linear
	in the message length.}

Thus motivated, after reviewing the GVW scheme
in Section~\ref{s:gvw} we introduce a
(non-universal) version of the lossy LZ scheme 
in \cite{kontoyiannis-lossy1-1}, termed LLZ,
and we compare its performance to that of 
GVW. Theorem~1 shows that LLZ is asymptotically 
optimal in the rate-distortion sense
for compressing data from a known discrete 
memoryless source with respect to a 
single-letter distortion criterion.
Simulation results 
are also presented, comparing
the performance of LLZ and GVW on a simple
Bernoulli source. These results indicate that
for blocklengths around 1000 bits,
GVW offers better compression than LLZ 
at a given distortion level, but 
it requires significantly more memory 
for its execution.
[The same findings are also confirmed in
the other simulation examples presented in Section~\ref{s:sim}.]

In order to combine the different advantages 
of the two schemes, in Section~\ref{s:hyb} we introduce
a hybrid algorithm (HYB), which 
utilizes both the divide-and-conquer 
idea of GVW and the single-database 
structure of LLZ. 
In Theorems~2 and~3 we prove that 
HYB shares with GVW the exact same 
rate-distortion 
performance and implementation complexity,
in that it operates in near-linear time
at the encoder and linear time at the
decoder. Moreover, like LLZ, 
the HYB scheme requires much less memory, 
by an unbounded factor, depending on the 
choice of parameters in the design 
of the two algorithms.
Experimental results 
are presented
in Section~\ref{s:sim}, comparing the performance 
of GVW and HYB.  These confirm the theoretical 
findings, and indicate that HYB outperforms 
existing schemes for the compression of some 
simple discrete sources with respect to the 
Hamming distortion criterion. The earlier
theoretical results stating that 
HYB's rate-distortion performance is 
the same as GVW's are confirmed empirically,
and it is also shown that, again for
blocklengths of approximately 1000 symbols,
the HYB scheme requires much less memory, 
by a factor ranging between
15 and 240.

After a brief discussion on 
potential extensions of the present results,
some conclusions are collected in 
Section~\ref{s:con}. The appendix contains the proofs
of the theorems in Sections~\ref{s:gvw}
and~\ref{s:hyb}.

% \newpage

\section{The GVW and LLZ algorithms}
\label{s:gvw}
%%%%%%%%%%%%%%%%%%%%%%%%%%%%%%%%%%%%%%%%%%%%%%%%%%%%%%%%%%%%%%%%%%

After describing the basic setting within which all
later results will be developed, in Section~\ref{s:gvw2}
we recall the divide-and-conquer idea of the
GVW scheme, and 
in Section~\ref{s:gvw3}
we present a new, non-universal lossy LZ algorithm
and examine its properties.

\subsection{The setting}
\label{s:gvw1}
%%%%%%%%%%%%%%%%%%%%%%%%%%%%%%%%%%%%%%%%%%%%%%%%%%%%%%%%%%%%%%%%%%

Let $\{X_n\}=\{X_1,X_2,\ldots\}$ be a memoryless
source on some finite alphabet $A$ and suppose 
that its distribution is described by a known 
probability mass function $P$ on $A$. 
The objective is to compress $\{X_n\}$ with
respect to a sequence of single-letter distortion
criteria,
$$\rho_n(x_1^n,y_1^n)=\frac{1}{n}\sum_{i=1}^n\rho(x_1,y_1),
\;\;\;\;n\geq 1,$$ 
where $x_1^n=(x_1,x_2,\ldots,x_n)\in A^n$ 
is an arbitrary source string 
to be compressed,
$y_1^n = (y_1,y_2,\ldots,y_n)$ 
is a potential reproduction string 
taking values in a
finite reproduction alphabet $\Ahat$, 
and $\rho:A\times\Ahat\to[0,\infty)$ 
is an arbitrary distortion measure.
We make the customary assumption that
for any source letter $x$ there
is a reproduction letter $y$ with
zero distortion,
$$\max_{x\in A}\min_{y\in\Ahat}\rho(x,y)=0.$$

The best achievable rate at which data
from the source $\{X_n\}$ can be compressed
with distortion not exceeding $D\geq 0$ is
given by the {\em rate-distortion function}
\cite{shannon:59}\cite{berger:book}\cite{cover:book},
\be
R(D)
\;=\;
\inf_{W(y|x):\sum_{x,y\in A}P(x)W(y|x)\rho(x,y)\leq D}
I(X;Y),
\label{eq:RDdefn}
\ee
where $I(X;Y)$ denotes the mutual information
between a random variable $X$ with the same
distribution $P$ as the source and a random
variable $Y$ with conditional distribution
$W(\cdot|x)$ given $X=x$.\footnote{The mutual
information, rate-distortion function, and
all other standard information-theoretic
quantities here and throughout are expressed
in bits; all logarithms are taken to be 
in base 2, unless stated otherwise.}
Let
$\Dmax=\min_{y\in\Ahat}E_P[\rho(X,y)]$; in
order to avoid the trivial case where $R(D)$
is identically equal to zero, 
$\Dmax$ is assumed to be strictly positive.
It is well-known and easy to check that,
for all distortion values in the nontrivial
range $0<D<\Dmax$, there is a conditional
distribution $W^*(\cdot|\cdot)$ that 
achieves the infimum in (\ref{eq:RDdefn}),
and this induces a distribution $Q^*$ on
$\Ahat$ via $Q^*(y)=\sum_{x\in A}P(x)W^*(y|x)$,
for all $y\in\Ahat$. With a slight abuse
of terminology (as $Q^*$ may not be unique)
we refer to $Q^*$ as {\em the optimal
reproduction distribution at distortion 
level $D$}. 
Recall also the analogous definition of 
the {\em distortion-rate function} $D(R)$ of the 
source; cf.\
\cite{berger:book}\cite{cover:book}.

\subsection{The GVW algorithm}
\label{s:gvw2}
%%%%%%%%%%%%%%%%%%%%%%%%%%%%%%%%%%%%%%%%%%%%%%%%%%%%%%%%%%%%%%%%%%
The GVW algorithm\footnote{To be more precise,
	this is one of two closely related schemes
	discussed in \cite{gupta-et-al:pre:09};
	see the relevant comments in Section~\ref{s:hyb}.}
is a fixed-rate, variable-distortion code of
blocklength $n$ and target distortion $D\in(0,\Dmax)$.
It is described in terms 
of two parameters; 
a ``small'' $\gamma>0$, and
an integer $\ell$ so 
that $n=k\ell$.

Given the target distortion level $D$,
let $R=R(D)+\gamma$, and take,
\be
\Dbar
=R^{-1}\Big( R(D)+\gamma/2\Big)
=D\Big( R(D)+\gamma/2\Big)
\leq D.
\label{eq:DbarGVW}
\ee
First a fixed-rate code of blocklength $\ell$ 
and rate $R$ is created according to 
Shannon's classical 
random codebook construction. 
Letting 
$Q^*$ denote the optimal reproduction distribution 
at level $\Dbar$,
the codebook consists
of $\lfloor 2^{\ell R}\rfloor$ i.i.d. 
codewords of length $\ell$,
each generated i.i.d. from $Q^*$. 
Writing
$x_1^n=x_1^\ell * x_{\ell+1}^{2\ell}*
\cdots * x_{(k-1)\ell+1}^{k\ell}$,
as the concatenation of $k$ sub-blocks,
each sub-block 
is matched to its $\rho_\ell$-nearest
neighbor in the codebook, and it is 
described to the decoder using 
$\lceil\log\lfloor 2^{\ell R}\rfloor\rceil\approx \ell R$ 
bits
to describe the index of that nearest
neighbor in the codebook.

This code is used $k$ times, once
on each of the $k$ sub-blocks,
to produce corresponding reconstruction 
strings $y_{(i-1)\ell +1}^{i\ell}$, for $i=1,2,\ldots,k$.
The description of $x_1^n$
is the concatenation of the descriptions 
of the individual sub-blocks, and the 
reconstruction string itself is the 
concatenation of the 
corresponding reproduction blocks,
$y_1^n=y_1^\ell * y_{\ell+1}^{2\ell}*
\cdots * y_{(k-1)\ell+1}^{k\ell}$.
The overall description length of this code
is $k\lceil\log\lfloor 2^{\ell R}\rfloor\rceil
\leq k\ell R=nR$ bits,
so the (fixed) rate of this code is $\leq R$ bits/symbol,
and its (variable) distortion is $\rho_n(x_1^n,y_1^n)$.

\subsection{The lossy Lempel-Ziv algorithm LLZ}
\label{s:gvw3}
%%%%%%%%%%%%%%%%%%%%%%%%%%%%%%%%%%%%%%%%%%%%%%%%%%%%%%%%%%%%%%%%%%

The LLZ algorithm described here can be seen as
a simplified (in that it is non-universal) 
and modified (to facilitate the comparison
below) version of the algorithm in
\cite{kontoyiannis-lossy1-1}.
It is a fixed-distortion, variable-rate code 
of blocklength $n$, described in terms 
of three parameters; an integer blocklength $\ell\leq n$,
and ``small'' $\alpha,\gamma>0$.\footnote{Note 
that in \cite{kontoyiannis-lossy1-1} a 
fixed-rate, variable-distortion {\em universal}
code is also described, but we restrict
attention here to the conceptually
simpler fixed-distortion algorithm.}
The algorithm will be presented in 
a setting ``dual'' to that of the GVW algorithm,
in the sense that was described in the Introduction.
The main difference is that the source sting 
$x_1^n$ will be parsed into substrings of {\em variable}
length, not of fixed length $\ell$.

Given $n$ and a target distortion level $D$,
define 
$R=R(D)+\gamma$, take,
$$\Dbar
=R^{-1}\Big( R(D)-\gamma/2\Big)
=D\Big( R(D)-\gamma/2\Big)
\geq D,$$
and let
$Q^*$ denote the optimal
reproduction distribution at level $\Dbar$.
Then generate a single i.i.d.\ 
database $Y_1^m=(Y_1,Y_2,\ldots,Y_m)$
of length,
\be
m=m(\ell)=\lfloor2^{\ell R}\rfloor+\ell-1,
\label{eq:m}
\ee
and make it available to both the 
encoder and decoder.

The encoding algorithm is as follows: The encoder 
calculates the length of the longest match (up to 
$(1+\alpha)\ell$-many symbols) of an initial portion 
of the message $x_1^n$, within distortion $\Dbar$,
in the database. Let 
$L_{\ell,1}$ denote the length
of this longest match,
$$L_{\ell,1}=\max\{1\leq k\leq (1+\alpha)\ell\,:\,
\rho_k(x_1^k,Y_i^{i+k-1})\leq \Dbar\;
\mbox{for some}\;1\leq i\leq m-k+1\;
\}, $$
and let $Z^{(1)}=x_1^{L_{\ell,1}}$ denote the initial 
phrase of length $L_{\ell,1}$ in $x_1^n.$
Then the encoder describes to the decoder:
\begin{itemize}
\item[$(a)$] the length $L_{\ell,1}$; this takes
$\lceil\log((1+\alpha)\ell)\rceil$ bits;
% this can be given 
% using
% \be
% \lceil\log(\lfloor(1+\alpha)\ell\rfloor)\rceil  
% \;\;\;\mbox{bits}.
% \label{eq:elias}
% \ee
% where $\log^c(k):=\lceil\log(\max\{2,k\})\rceil$;
% cf. \cite{elias}\cite{wyner-ziv:2};
% \be
% \log^c(L_{m,1}) + 
% 2\log^c(\log^c(L_{m,1}))+2
% \;\;\;\mbox{bits},
% \label{eq:elias}
% \ee
% where $\log^c(k):=\lceil\log(\max\{2,k\})\rceil$;
% cf. \cite{elias}\cite{wyner-ziv:2};

\item[$(b)$] the position $i$ in the database where the
match occurs; this takes
$\lceil\log m\rceil$ bits.
\end{itemize}
From $(a)$ and $(b)$ the decoder can 
recover the string 
$\hat{Z}^{(1)}\,=\,Y_i^{i+L_{\ell,1}-1},$
which is within distortion $\Dbar$ of 
$Z^{(1)}$.

% The description of the match-length $L_{\ell,1}$ in $(a)$
% is given in $B(L_{m,1})$ bits, where $B(\cdot)$ is
% the length-function of a prefix code defined as,
% \be
% B(x)=\lceil\beta\log x +C\rceil,
% \label{eq:gauss}
% \ee
% where $C$ is chosen so that Kraft's inequality
% is satisfied,
% $$
% C:=
% \log\left(
% \sum_{x=1}^{\lfloor(1+\alpha)\ell\rfloor} 2^{-\beta\log x}
% \right)
% =\log\left(
% \sum_{x=1}^{\lfloor(1+\alpha)\ell\rfloor} x^{-\beta}
% \right).
% $$
% The description length of 
% $(a)$ and $(b)$ is bounded above by
% \be
% \log(L_{m,1}+1) +
% 2\log\log(L_{m,1}+1) +
% + \log m + 2
% \;\;\;\;\mbox{bits.}
% \label{eq:desc1}
% \ee

Alternatively, $Z^{(1)}$ can be described 
with {\em zero} distortion by first describing 
its length $L_{\ell,1}$ as before,
and then describing
$Z^{(1)}$ itself
% ({\em not} $\hat{Z}^{(1)}$!) 
directly using,
\be
\lceil L_{\ell,1}\log |\Ahat|\rceil\;\;\;\;\mbox{bits.}
\label{eq:desc2}
\ee
The encoder uses whichever one of the two descriptions is
shorter. [Note that is not necessary to add a flag to 
indicate which one was chosen; the decoder can simply
check if 
$\lceil L_{\ell,1}\log |\Ahat|\rceil$
is larger or smaller than $\lceil\log m\rceil$.]
Therefore, from~$(a)$, $(b)$, and~(\ref{eq:desc2})
the length of the description of
$Z^{(1)}$ is,
\be
% \lceil\log(\lfloor(1+\alpha)\ell\rfloor)\rceil  
\lceil\log((1+\alpha)\ell)\rceil
+\min\{
\lceil\log m\rceil,
\,\lceil L_{\ell,1}\log |\Ahat|\rceil\}
\;\;\;\mbox{bits}.
\label{eq:desc4}
\ee
% bounded above by 
% \be
% \min \left\{
                % C_1\log(L_{m,1}+1) +
                % \log m,
                % \;C_3L_{m,1}
        % \right\}\;\;\;\;\mbox{bits,}
% \label{eq:desc3}
% \ee
% for some fixed constants $C_1,\,C_3,$ 
% independent of $m,$ $n$, and $x_1^n$.

After $Z^{(1)}$ has been described within
distortion $\Dbar$, the same process is repeated 
to encode the rest of the message:
The encoder finds the length $L_{\ell,2}$ of the longest 
string starting at position $(L_{\ell,1}+1)$ in $x_1^n$ that
matches within distortion $\Dbar$ into the
database, and describes 
$Z^{(2)}=x_{L_{\ell,1}+1}^{L_{\ell,1}+L_{\ell,2}}$
to the decoder by repeating the above steps. 
The algorithm is terminated, in the natural way, when the
entire string $x_1^n$ has been exhausted. At that point,
$x_1^n$ has been parsed into $\Pi_\ell=\Pi_\ell(x_1^n,D)$ 
distinct phrases $Z^{(k)}$, each of length $L_{\ell,k}$,
$x_1^n\;=\;Z^{(1)}*Z^{(2)}*\cdots *Z^{(\Pi_\ell)},$
with the possible exception of the last phrase,
which may be shorter. Since each substring $Z^{(k)}$ is 
described within distortion $\Dbar$, also the 
concatenation of all the reproduction strings, 
call it $\psi_1^n:=\hat{Z}^{(1)}*\hat{Z}^{(2)}*\cdots* \hat{Z}^{(\Pi_\ell)}$,
will be within distortion $\Dbar$ of $x_1^n$.

The distortion achieved by this code is
$\rho_n(x_1^n,\psi_1^n)$, and it is
guaranteed to be $\leq \Dbar $ by construction.
Regarding the rate,
if we write $\Lambda(x_1^n)=\Lambda(x_1^n,\ell,D)$ for the 
overall description length of $x_1^n$,
then from (\ref{eq:desc4}),
\be
\Lambda(x_1^n)=
      \sum_{k=1}^{\Pi_\ell}
	\Big[
\lceil\log((1+\alpha)\ell)\rceil  
+\min\{
\lceil\log m\rceil,
\,\lceil L_{\ell,k}\log |\Ahat|\rceil\}
	\Big]
\;\;\mbox{bits,}
\label{eq:ell-bound}
\ee
and the rate achieved by this code is $\Lambda(x_1^n)/n$ 
bits/symbol.

\medskip

\noindent
{\bf Remark. }
As mentioned in the Introduction, there are
two main differences between the GVW algorithm
and the LLZ scheme. The first one is that while
the GVW is based on a Shannon-style random codebook,
the LLZ uses an LZ-type random database. The second
is that GVW divides up the message $x_1^n$ into fixed-length 
sub-blocks of size $\ell$, whereas LLZ parses $x_1^n$
into variable-length strings of (random) 
lengths $L_{\ell,k}$. But there is also an important
point of solidarity between the two algorithms.
Recall
\cite[Theorem~23]{dembo-kontoyiannis:wyner} 
that, for large $\ell$, the match length
$L_{\ell,1}$ behaves logarithmically in the
size of the database; that is, with high
probability,
$$L_{\ell,1}\approx\frac{\log m(\ell)}{R(\Dbar)}\approx\ell,$$
where the second approximation follows by the
choice of $m(\ell)$ and of $\Dbar.$
Therefore, both algorithms end up parsing the 
message $x_1^n$ into sub-blocks of length
$\approx\ell$ symbols.

\medskip

Our first result shows that LLZ is asymptotically
optimal in the usual sense established for 
fixed-database versions of LZ-like schemes;
see \cite{wyner-ziv:3}\cite{kontoyiannis-lossy1-1}.
Specifically, it is shown that by taking $\ell$
large enough and $\gamma$ small enough, the
LLZ comes arbitrarily close to any optimal
rate-distortion point $(R(D),D)$.
Note that $\alpha>0$ is a parameter that
simply controls the complexity of the 
best-match search, and its influence
on the rate-distortion performance is
asymptotically irrelevant.

\medskip

\noindent
{\bf Theorem 1. }{\sc [LLZ Optimality] } 
Suppose the LLZ with parameters
$\ell,\alpha$ and $\gamma$ is used
to compress a memoryless source $\{X_n\}$
with rate-distortion function $R(D)$
at a target distortion rate $D\in(0,\Dmax)$.
For any $\delta>0$, 
the parameter $\gamma>0$ 
can be chosen small enough
such that:\\
(a) For any choice of $\ell$ and
any blocklength $n,$
the distortion achieved by LLZ 
is no greater than
$D+\delta.$\\
(b) Taking $\ell$ large enough,
the asymptotic rate of LLZ achieves
the rate-distortion bound, in that,
\be
\limsup_{\ell\to\infty}\limsup_{n\to\infty}\;
E\left\{\frac{\Lambda(X_1^n,\ell,D)}{n}\;\Big|\,X_1^n\right\}
\;\leq\; R(\Dbar) = R(D)-\gamma/2
\;\;\;\;\mbox{bits/symbol, w.p.1},
\label{eq:as-optimality}
\ee
where the expectation is over all databases.
Therefore, also,
\be
\limsup_{\ell\to\infty}\limsup_{n\to\infty}\;
E\left\{\frac{\Lambda(X_1^n,\ell,D)}{n}\right\}\;\leq\; 
R(\Dbar) = R(D)-\gamma/2
\;\;\;\;\mbox{bits/symbol,}
\label{eq:optimality}
\ee
with the expectation here being 
over both the message 
and the databases.

\medskip

Next, the performance of LLZ is compared
with that of GVW on data generated
from a Bernoulli source with parameter
$p=0.4$ and with respect to Hamming distortion.
Simulation results at different target 
distortions are shown in 
Figure~\ref{f:Bern04a} and 
Table~\ref{tab:04a}; see Section~\ref{s:sim} for
details on the choice of parameter values.
It is clear from these results that,
at the same distortion level, the GVW
algorithm typically gives a better rate 
than LLZ.
In terms of implementation complexity,
the two algorithms have comparable
execution times, but the LLZ uses 
significantly less memory.
The same pattern -- GVW giving better
compression but using much more memory
than LLZ -- is also confirmed in the 
other examples we consider in Section~\ref{s:sim}.

Note that, like for the case of GVW,
more can be said about the implementation
complexity of LLZ and how it depends
on the exact choice of parameters
$\ell,\alpha$ and $\gamma$. But since,
as we will see next, the performance
of both algorithms is dominated by that 
of a different algorithm (HYB), 
we do not pursue this direction
further.

\begin{figure}[ht!]
\centerline{\includegraphics[width=3.6in,height=3.4in]{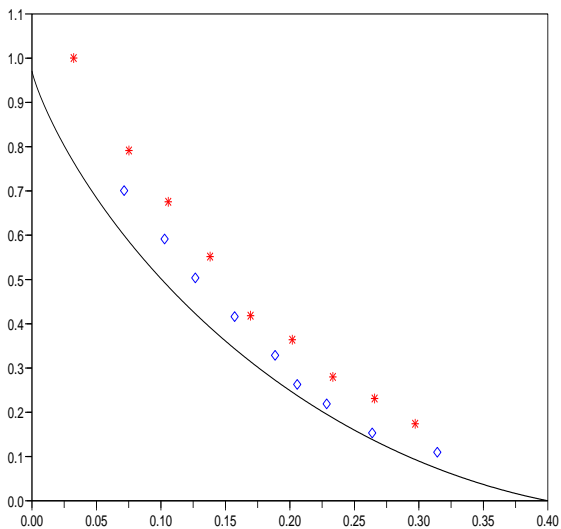}}
     \caption{Comparison of the rate-distortion
	performance of LLZ vs.\ that of GVW,
	on a data string of length
	$n=1050$ bits generated from a Bernoulli
	source with parameter
	$p=0.4$. The solid line is the rate-distortion
	function, the rate-distortion pairs achieved
	by LLZ are shown as red stars and those of
	GVW as blue diamonds.
	}
\flabel{Bern04a}
\end{figure}

\begin{table}[ht!]
  \begin{center}
    \begin{tabular}{|c||c|c|c|c|c|}
      \hline
      \multicolumn{6}{|c|}{{\bf Bern(0.4) source, Hamming distortion}}\\
      \hline
      & \multicolumn{5}{|c|}{\em Performance parameters}\\
        \hline
{\em Algorithm}& 
	$D_{\rm target}$&
	$D_{\rm achieved}$& 
	rate&
	memory&
		time \\
     \hline
GVW & 0.05 & 0.07143 & 0.70095 & 26MB & 27m53s \\
	\hline
GVW & 0.08 & 0.10286 & 0.59143 & 23MB & 21m11s \\
	\hline
GVW & 0.11 & 0.12667 & 0.50381 & 27MB & 20m48s \\
	\hline
GVW & 0.14 & 0.15714 & 0.41619 & 31MB & 19m52s \\
	\hline
GVW & 0.17 & 0.18857 & 0.32857 & 36MB & 18m48s \\
	\hline
GVW & 0.2 & 0.20571 & 0.26286 & 46MB & 19m18s \\
	\hline
GVW & 0.23 & 0.22857 & 0.21905 & 57MB & 18m42s \\
	\hline
GVW & 0.26 & 0.26381 & 0.15333 & 79MB & 19m46s \\
	\hline
GVW & 0.29 & 0.31429 & 0.10952 & 113MB & 20m29s \\
	\hline
	\hline
LLZ & 0.05 & 0.03238 & 1.00029 & 1.5MB & 4m23s \\
	\hline
LLZ & 0.08 & 0.07524 & 0.79129 & 1.28MB & 6m15s \\
	\hline
LLZ & 0.11 & 0.10571 & 0.6754 & 1.46MB & 8m53s \\
	\hline
LLZ & 0.14 & 0.1381 & 0.55171 & 1.69MB & 11m18s \\
	\hline
LLZ & 0.17 & 0.16952 & 0.41827 & 2.6MB & 18m15s \\
	\hline
LLZ & 0.2  & 0.2019 & 0.36381 & 3.6MB & 20m09s \\
	\hline
LLZ & 0.23 & 0.23333 & 0.27975 & 6.2MB & 41m32s \\
	\hline
LLZ & 0.26 & 0.26571 & 0.23102 & 13MB & 63m56s \\
	\hline
LLZ & 0.29 & 0.29714 & 0.1741 & 47MB & 165m54s \\
     \hline
     \end{tabular}
  \end{center}
     \caption{Comparison of the performance
	of LLZ vs.\ that of GVW
	on a data string of length
	$n=1050$ bits generated from a Bernoulli
	source with parameter $p=0.4$. 
	}
     \label{tab:04a}
\end{table}

\newpage

$\;$

\newpage

\section{The HYB algorithm}
\label{s:hyb}
%%%%%%%%%%%%%%%%%%%%%%%%%%%%%%%%%%%%%%%%%%%%%%%%%%%%%%%%%%%%%%%%%%

In order to combine the rate-distortion advantage
of GVW with the memory advantage of LLZ, in this 
section we introduce a hybrid algorithm and
examine its performance.

The new algorithm, termed HYB, uses the divide-and-conquer
approach of GVW, but based on a random database
like the LLZ instead of a
random codebook. It is
a fixed-rate, variable-distortion code of
blocklength $n$ and target distortion $D\in(0,\Dmax)$,
and it is described in terms 
of two parameters; 
a ``small'' $\gamma>0$, and
an integer $\ell$ so that $n=k\ell$.

Like with the GVW, given a target distortion 
level $D$, let $R=R(D)+\gamma$ and take
$\Dbar$ as in (\ref{eq:DbarGVW}).
Now, like for the LLZ algorithm,
let $m=m(\ell)=\lfloor2^{\ell R}\rfloor+\ell-1$
as in (\ref{eq:m}),
and generate a random 
database $Y_1^m=(Y_1,Y_2,\ldots,Y_m)$,
where the $Y_i$ are drawn i.i.d.\
from the optimal reproduction 
distribution at level $\Dbar$.
The database is made available to both
the encoder and the decoder,
and the message $x_1^n$ to be
compressed is parsed into
$k=n/\ell$ non-overlapping blocks,
$x_1^n=x_1^\ell * x_{\ell+1}^{2\ell}*
\cdots * x_{(k-1)\ell+1}^{k\ell}$.

The first sub-block $x_1^\ell$ 
is matched to its $\rho_\ell$-nearest
neighbor in the database,
where we consider each possible
$Y_i^{i+\ell-l}$,
$i=1,2,\ldots,\lfloor 2^{\ell R}\rfloor$
as a potential
reproduction word.
Then $x_1^\ell$
is described to the decoder 
by describing the position of
its matching reproduction block
in the database using $\approx\ell R$ bits,
and the same process is repeated 
on each of the $k$ sub-blocks,
to produce $k$ reconstruction 
strings.
The description of $x_1^n$ is the 
concatenation of the descriptions 
of the individual sub-blocks, 
and the reconstruction string itself 
is the concatenation of the 
corresponding reproduction blocks.
The overall description length of this code
is $k\lceil\log\lfloor 2^{\ell R}\rfloor\rceil
\leq k\ell R=nR$ bits.

The following result shows that the HYB algorithm shares
the exact same rate-distortion performance,
as well as the same implementation complexity
characteristics,
as the GVW. Let:
$$\hat{\gamma}=\min\{1,2(R(D/2)-R(D))\}.$$

\medskip

\noindent
{\bf Theorem 2. }{\sc [HYB Compression/Complexity Trade-off] }
Consider a memoryless source $\{X_n\}$
with rate-distortion function $R(D)$, which is
to be compressed at target distortion
level $D\in(0,\Dmax)$. There exists an
$\hat{\epsilon}>0$ such that, 
for any $0<\epsilon<\hat{\epsilon}$,
the HYB algorithm with parameters
$0<\gamma<\hat{\gamma}$ and $\ell$
as in (\ref{eq:ellDefn1})
achieves a rate of $R=R(D)+\gamma$ bits/symbol, 
its expected distortion is less 
than $D+\epsilon$, and moreover:

-- Encoding time per source symbol is proportional
to $(\lambda_1/\epsilon)^{\lambda_2(D)/\gamma^2}$,

-- Decoding time per symbol is independent
of $\gamma$ and $\epsilon$,

\noindent
where $\lambda_1$ and $\lambda_2(D)$ are independent
of $\epsilon$ and $\gamma$.

\medskip

\noindent
{\bf Remarks. }

1. Theorem~2 is an exact analog of Theorem~1 proved
for GVW in \cite{gupta-et-al:pre:09}, the only difference
being that we consider average distortion instead
of the probability-of-excess distortion criterion. The
reason is that, instead of presenting an existence
proof for an algorithm with certain desired properties,
here we examine the performance of the HYB algorithm itself. 
Indeed, the proof
of Theorem~2 can easily be modified to prove the stronger
claim that there exists {\em some} instance of the random
database $Y_1^m$ such that, using that particular database, 
the HYB algorithm also has the additional property that
the probability of excess distortion vanishes as $n\to\infty$.
The same comments apply to Theorem 3 below.

2. In \cite{gupta-et-al:pre:09} a similar result is proved
with the roles and $\epsilon$ and $\gamma$ interchanged.
In fact, it should be pointed out that the scheme
we call ``the'' GVW algorithm here corresponds to 
the scheme used in the proof of 
\cite[Theorem~1]{gupta-et-al:pre:09}. A slight variant
(having to do with the choice of parameter values and
not with the mechanics of the algorithm itself) is used
to prove \cite[Theorem~2]{gupta-et-al:pre:09}. Having gone
over the proofs, it would be obvious to the reader that, 
once the corresponding changes are made for HYB,
an analogous result 
can be proved for HYB. The straightforward
but tedious details are omitted.

3. In terms of memory, the GVW scheme requires
$\ell\lfloor 2^{\ell R}\rfloor$
reproduction symbols for storing the codebook,
while using the same memory parameters
the HYB algorithm needs 
$m(\ell)=\lfloor2^{\ell R}\rfloor+\ell-1$
symbols. The ratio between the two
is,
$$\frac{\mbox{memory for GVW}}
{\mbox{memory for HYB}}
=
\frac{\ell\lfloor 2^{\ell R}\rfloor}
{\lfloor2^{\ell R}\rfloor+\ell-1}
\approx\ell,$$
so that the GVW needs $\approx\ell$ 
{\em times}
more memory than HYB. Moreover,
the closer we require the algorithm
to come to achieving an optimal
$(D,R(D))$ point, the smaller the
values of $\epsilon$ and $\alpha$
need to be taken in Theorem~2,
and the larger the corresponding
value of $\ell$; cf. equation (\ref{eq:ellDefn1}). 
Therefore, not
only the difference, but even the
ratio of the memory required by
GVW compared to HYB, is unbounded.

\medskip

The next result shows that, choosing the parameters
$\ell$ and $\gamma$ in HYB appropriately, 
optimal compression performance can be 
achieved with linear decoding complexity and 
near-linear encoding complexity.
It is a parallel result 
to \cite[Theorem 3]{gupta-et-al:pre:09}.

\medskip

\noindent
{\bf Theorem 3. }{\sc [HYB Near-Linear Complexity] }
For a memoryless source $\{X_n\}$ with rate-distortion
function $R(D)$, a target distortion level $D\in(0,\Dmax)$,
and an arbitrary increasing and unbounded function 
$g(n)$, the HYB algorithm with appropriately
chosen parameters $\ell=\ell(n)$ and $\gamma=\gamma(n)$,
achieves a limiting rate equal to
$R(D)$ bits/symbol and limiting average
distortion $D$. The encoding and decoding 
complexities are $O(ng(n))$ and $O(n)$ respectively.

\medskip

The actual empirical performance of HYB on simulated
data is compared to that of GVW and LLZ in the following 
section.

\newpage 

\section{Simulation results}
\label{s:sim}
%%%%%%%%%%%%%%%%%%%%%%%%%%%%%%%%%%%%%%%%%%%%%%%%%%%%%%%%%%%%%%%%%%

Here the empirical performance of the
HYB scheme is compared with that of GVW and LLZ, on three
simulated data sets from
simple memoryless sources.\footnote{We do not present
	comparison results with earlier schemes 
	apart from the GVW, since extensive such studies already 
	exist in the literature; in particular, the GVW is compared
	in \cite{gupta-et-al:pre:09} with the algorithms
	proposed in \cite{maneva-wainwright:05},
	\cite{gupta-verdu:pre} and \cite{rissanen-tabus}.
	} The following 
parameter values were used
in all of the experiments.
For the GVW and HYB algorithms, 
$\ell$ was chosen as in \cite{gupta-et-al:pre:09} to be
$\ell =\lceil\frac{22}{R(D)+\gamma}\rceil$,
where $R(D)$ is the rate-distortion function of the
source, and $\gamma$ was taken 
equal to
$0.002$. Similarly, for LLZ we took
$\ell =\lceil 22/R(D)\rceil$, $\gamma=0.03$ and $\alpha=0.1$.
Note that, with this choice of parameters, the complexity
of all three algorithms is approximately linear
in the message length $n$.
All experiments were performed on a Sony Vaio laptop
running Ubuntu Linux, under identical 
conditions.\footnote{Although there is 
	a wealth of efficient algorithms for 
	the problem of approximate string
	matching (see, e.g., \cite{crochemore:book}\cite{
	arnaud-szpankowski}\cite{atallah-et-al}\cite{crochemore:1}
	and the references therein),
	since HYB clearly outperforms LLZ,
	our version of the LLZ 
	scheme was implemented using the naive,
	greedy scheme consistent with 
	the definition of algorithm.}

First we revisit the example of Section~\ref{s:gvw}; 
$n=1050$ bits generated by a Bernoulli source 
with parameter $p=0.4$, are compressed by all 
three algorithms at various different distortion
levels with respect to Hamming
distortion. Figure~\ref{f:Bern04} shows the
rate-distortion pairs achieved.

\begin{figure}[ht!]
\centerline{\includegraphics[width=4.2in,height=3.8in]{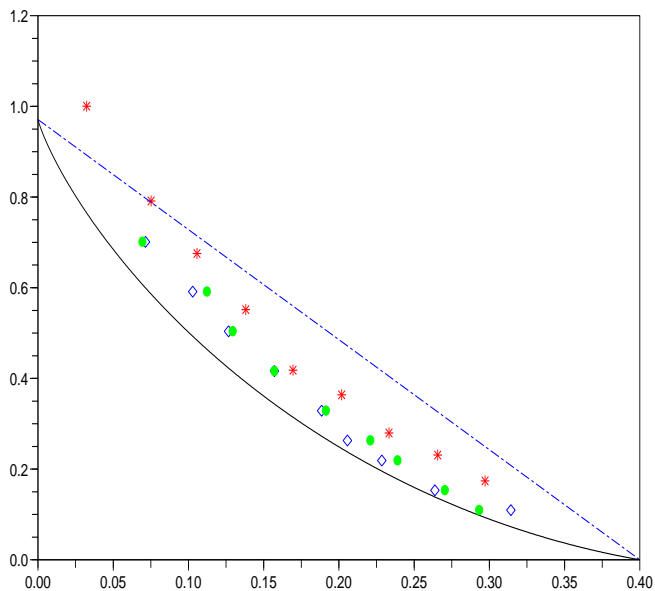}}
     \caption{Comparison of the rate-distortion
	performance of GVW, LLZ and HYB
	on a data string of length
	$n=1050$ bits generated from a Bernoulli
	source with parameter
	$p=0.4$. The solid convex curve is the rate-distortion
	function; the rate-distortion pairs achieved
	by GVW are shown as blue diamonds;
	by LLZ as red stars; and by HYB as
	bold green dots.
	}
\flabel{Bern04}
\end{figure}

\noindent
{\bf Rate-distortion performance. }
It is evident that the compression performance obtained 
by GVW and HYB is near-identical, and better than that of LLZ.
This example was also examined by Rissanen and Tabus
in \cite{rissanen-tabus}, where it was noted that it
is quite hard for any implementable scheme to produce
rate-distortion pairs below the straight line connecting
the end-points $(D,R(D))$ of the rate-distortion curve 
corresponding to $D=0$ and $D=0.4$. 
As noted in \cite{gupta-et-al:pre:09},
the Rissanen-Tabus scheme produces results slightly
below the straight line, and it is one of 
the best implementable schemes for this problem.

\begin{table}[ht!]
  \begin{center}
    \begin{tabular}{|c||c|c|c|c|c|}
      \hline
      \multicolumn{6}{|c|}{{\bf Bern(0.4) source, Hamming distortion}}\\
      \hline
      & \multicolumn{5}{|c|}{\em Performance parameters}\\
        \hline
{\em Algorithm}& 
	$D_{\rm target}$&
	$D_{\rm achieved}$& 
	rate&
	memory&
		time \\
     \hline
HYB & 0.05 & 0.06952 & 0.70095 & 0.79MB & 2m45s \\
	\hline
HYB & 0.08 & 0.11238 & 0.59143 & 0.6MB & 3m06s \\
	\hline
HYB & 0.11 & 0.12952 & 0.50381 & 0.59MB & 3m33s \\
	\hline
HYB & 0.14 & 0.15714 & 0.41619 & 0.56MB & 4m06s \\
	\hline
HYB & 0.17 & 0.19143 & 0.32857 & 0.52MB & 4m40s \\
	\hline
HYB & 0.2 & 0.22095 & 0.26286 & 0.53MB & 5m21s \\
	\hline
HYB & 0.23 & 0.23905 & 0.21905 & 0.51MB & 5m26s \\
	\hline
HYB & 0.26 & 0.27048 & 0.15333 & 0.53MB & 6m27s \\
	\hline
HYB & 0.29 & 0.29333 & 0.10952 & 0.53MB & 6m56s \\
     \hline
     \end{tabular}
  \end{center}
     \caption{Performance 
	achieved by the 
	HYB algorithm
	on a data string of length
	$n=1050$ bits generated from a Bernoulli
	source with parameter $p=0.4$. 
	}
     \label{tab:04}
\end{table}

\noindent
{\bf Memory and complexity. }
Tables~\ref{tab:04a} and~\ref{tab:04} contain a 
complete listing off all 
performance parameters obtained in the above experiment,
including the execution time required for the encoder
and the total amount of memory used.
As already observed in Section~\ref{s:gvw},
the LLZ scheme requires much less 
memory that GVW, and so does
the hybrid algorithm HYB.
In fact, while GVW and HYB produce
essentially identical rate-distortion
performance, the HYB algorithm requires
between 32 and 213 {\em times} less memory
than GVW. [Note that these figures
are deterministic; the memory requirement
is fixed by the description of the algorithm
and it is not subject to random variations
produced by the simulated data.]
In terms of the corresponding execution times,
the GVW and HYB share the exact same
theoretical complexity in their implementation.
Nevertheless, because of the vastly different
memory requirements, in practice we find that
the execution times of HYB were approximately
three to ten times faster than GVW.

\medskip

The second example is again
on a Bernoulli source with respect
to Hamming distortion, this time with 
source parameter $p=0.2$.
The corresponding simulation 
results 
are displayed in 
Figure~\ref{f:Bern02} 
and Table~\ref{tab:02}.

Finally, in the third example 
$\{X_n\}$ is taken as a memoryless 
source uniformly distributed on $\{0,1,2,3\}$,
to be compressed with respect to 
Hamming distortion. The empirical
results are shown in 
Figure~\ref{f:4letter} 
and Table~\ref{tab:4letter}.

In both these cases, the same 
qualitative conclusions are drawn.
The rate-distortion
performance of the GVW and HYB algorithms
is essentially indistinguishable, while
the compression achieved by LLZ is 
generally somewhat worse, though in several 
instances not significantly so.
In the second example note that 
the memory required by HYB is smaller than
that of GVW by a factor that ranges between
44 and 242, while in the third example
the corresponding factors are between
16 and 218.
And again, although the theoretical 
implementation complexity of GVW and HYB
is identical, because of their different
memory requirements the encoding time
of HYB is smaller than that of GVW by
a factor ranging between approximately
3 and 9 in the second example,
and between 1.25 and 1.5 in the third
example.

\begin{figure}[ht!]
\centerline{\includegraphics[width=4.2in,height=3.5in]{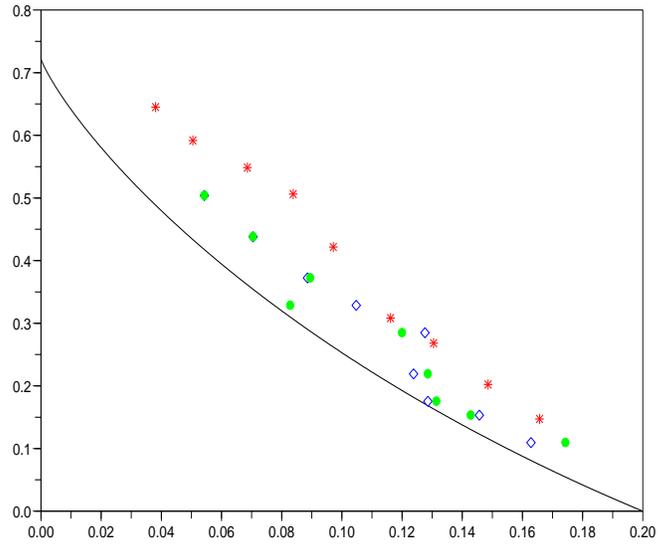}}
     \caption{Comparison of the rate-distortion
	performance of GVW, LLZ and HYB
	on a data string of length
	$n=1050$ bits generated from a Bernoulli
	source with parameter
	$p=0.2$. The solid curve is the rate-distortion
	function; the rate-distortion pairs achieved
	by GVW are shown as blue diamonds;
	by LLZ as red stars; and by HYB as
	bold green dots.
	}
\flabel{Bern02}
\end{figure}

\begin{table}[ht!]
  \begin{center}
    \begin{tabular}{|c||c|c|c|c|c|}
      \hline
      \multicolumn{6}{|c|}{{\bf Bern(0.2) source, Hamming distortion}}\\
      \hline
      & \multicolumn{5}{|c|}{\em Performance parameters}\\
        \hline
{\em Algorithm}& 
	$D_{\rm target}$&
	$D_{\rm achieved}$& 
	rate&
	memory&
		time \\
	\hline
GVW & 0.04  & 0.05429 & 0.50381 & 25MB & 19m05s \\
	\hline
GVW & 0.055 & 0.07048 & 0.4381 & 28MB & 18m13s \\
	\hline
GVW & 0.07  & 0.08857 & 0.37238 & 35MB & 19m50s \\
	\hline
GVW & 0.085 & 0.10476 & 0.32857 & 42MB & 20m14s \\
	\hline
GVW & 0.1   & 0.12762 & 0.28476 & 49MB & 19m55s \\
	\hline
GVW & 0.115 & 0.12381 & 0.21905 & 59MB & 20m03s \\
	\hline
GVW & 0.13 & 0.12857 & 0.17524 & 73MB & 19m57s \\
	\hline
GVW & 0.145 & 0.14571 & 0.15333 & 90MB & 19m08s \\
	\hline
GVW & 0.16 & 0.16286 & 0.10952 & 126MB & 19m38s \\
	\hline
	\hline
LLZ & 0.04  & 0.0381 & 0.64495 & 1.36MB & 3m05s \\
	\hline
LLZ & 0.055 & 0.05048 & 0.59165 & 2.02MB & 7m45s \\
	\hline
LLZ & 0.07  & 0.06857 & 0.54836 & 1.9MB & 8m02s \\
	\hline
LLZ & 0.085 & 0.08381 & 0.50616 & 2.4MB & 13m38s \\
	\hline
LLZ & 0.1   & 0.09714 & 0.42154 & 3.1MB & 22m18s \\
	\hline
LLZ & 0.115 & 0.11619 & 0.3083 & 5.2MB & 24m03s \\
	\hline
LLZ & 0.13  & 0.13048 & 0.26809 & 8.3MB & 58m07s \\
	\hline
LLZ & 0.145 & 0.14857 & 0.20223 & 21MB & 132m30s \\
	\hline
LLZ & 0.16  & 0.16571 & 0.1472 & 100MB & 377m10s \\
	\hline
	\hline
HYB    & 0.04  & 0.05429 & 0.50381 & 0.56MB & 2m02s \\
	\hline
HYB    & 0.055 & 0.07048 & 0.4381 & 0.53MB & 2m54s \\
	\hline
HYB    & 0.07  & 0.08952 & 0.37238 & 0.57MB & 3m32s \\
	\hline
HYB    & 0.085 & 0.08286 & 0.32857 & 0.58MB & 3m52s \\
	\hline
HYB    & 0.1   & 0.12 & 0.28476 & 0.57MB & 4m46s \\
	\hline
HYB    & 0.115 & 0.12857 & 0.21905 & 0.56MB & 5m21s \\
	\hline
HYB    & 0.13  & 0.13143 & 0.17524 & 0.55MB & 5m45s \\
	\hline
HYB    & 0.145 & 0.14286 & 0.15333 & 0.52MB & 6m30s \\
	\hline
HYB    & 0.16  & 0.17429 & 0.10952 & 0.52MB & 7m11s \\
     \hline
     \end{tabular}
  \end{center}
     \caption{Comparison of the performance
	of GVW, LLZ and HYB
	on a data string of length
	$n=1050$ bits generated from a Bernoulli
	source with parameter $p=0.2$. 
	}
     \label{tab:02}
\end{table}

\newpage

\begin{figure}[ht!]
\centerline{\includegraphics[width=4.2in,height=3.5in]{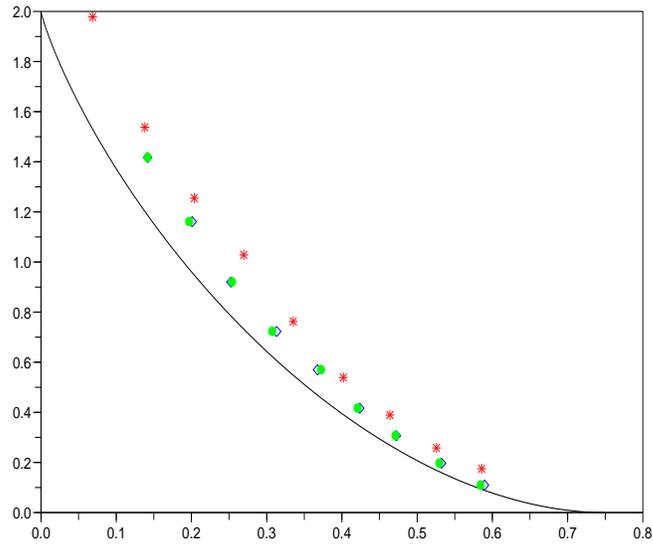}}
     \caption{Comparison of the rate-distortion
	performance of GVW, LLZ and HYB
	on a data string of length
	$n=1050$ symbols generated from 
	the Uniform source on $\{0,1,2,3\}$.
	The solid curve is the rate-distortion
	function; the rate-distortion pairs achieved
	by GVW are shown as blue diamonds;
	by LLZ as red stars; and by HYB as
	bold green dots.
	}
\flabel{4letter}
\end{figure}

\begin{table}[ht!]
  \begin{center}
    \begin{tabular}{|c||c|c|c|c|c|}
      \hline
      \multicolumn{6}{|c|}{{\bf $U\{0,1,2,3\}$ source, Hamming distortion}}\\
      \hline
      & \multicolumn{5}{|c|}{\em Performance parameters}\\
        \hline
{\em Algorithm}& 
	$D_{\rm target}$&
	$D_{\rm achieved}$& 
	rate&
	memory&
		time \\
	\hline
GVW & 0.1  & 0.1419 & 1.41714 & 43MB & 10m27s \\
	\hline
GVW & 0.16 & 0.20095 & 1.16095 & 24MB & 6m44s \\
	\hline
GVW & 0.22 & 0.25238 & 0.92 & 31MB & 8m19s \\
	\hline
GVW & 0.28 & 0.31333 & 0.72286 & 44MB & 11m12s \\
	\hline
GVW & 0.34 & 0.36762 & 0.56952 & 43MB & 9m45s \\
	\hline
GVW & 0.4  & 0.42381 & 0.41619 & 65MB & 12m29s \\
	\hline
GVW & 0.46 & 0.47238 & 0.30667 & 92MB & 13m59s \\
	\hline
GVW & 0.52 & 0.53238 & 0.19714 & 124MB & 14m12s \\
	\hline
GVW & 0.58 & 0.58952 & 0.10952 & 229MB & 17m30s \\
	\hline
	\hline
LLZ & 0.1  & 0.06857 & 1.97778 & 3.597MB & 9m54s \\
	\hline
LLZ & 0.16 & 0.1381 & 1.53794 & 1.79MB & 7m46s \\
	\hline
LLZ & 0.22 & 0.20381 & 1.25461 & 2.04MB & 12m50s \\
	\hline
LLZ & 0.28 & 0.26952 & 1.02841 & 2.61MB & 18m51s \\
	\hline
LLZ & 0.34 & 0.33524 & 0.76228 & 3.445MB & 28m25s \\
	\hline
LLZ & 0.4  & 0.4019 & 0.5393 & 3.49MB & 30m37s \\
	\hline
LLZ & 0.46 & 0.46381 & 0.3893 & 5.44MB & 46m19s \\
	\hline
LLZ & 0.52 & 0.52571 & 0.25807 & 14.6MB & 105m56s \\
	\hline
LLZ & 0.58 & 0.58571 & 0.17475 & 104MB & 62m16s \\
	\hline
	\hline
HYB    & 0.1  & 0.1419 & 1.41714 & 2.58MB & 7m49s \\
	\hline
HYB    & 0.16 & 0.19714 & 1.16095 & 1.22MB & 5m06s \\
	\hline
HYB    & 0.22 & 0.25429 & 0.92 & 1.26MB & 6m37s \\
	\hline
HYB    & 0.28 & 0.30762 & 0.72286 & 1.39MB & 8m48s \\
	\hline
HYB    & 0.34 & 0.37238 & 0.56952 & 1.05MB & 7m42s \\
	\hline
HYB    & 0.4  & 0.42095 & 0.41619 & 1.18MB & 9m39s \\
	\hline
HYB    & 0.46 & 0.47143 & 0.30667 & 1.15MB & 10m34s \\
	\hline
HYB    & 0.52 & 0.52952 & 0.19714 & 1.01MB & 10m14s \\
	\hline
HYB    & 0.58 & 0.58476 & 0.10952 & 1.05MB & 11m43s \\
     \hline
     \end{tabular}
  \end{center}
     \caption{Comparison of the performance
	of GVW, LLZ and HYB
	on a data string of length
	$n=1050$ symbols generated from 
	the Uniform source on $\{0,1,2,3\}$.
	}
     \label{tab:4letter}
\end{table}

\newpage 

$\;$

\newpage 

$\;$

\newpage 

\section{Conclusions and Extensions}
\label{s:con}
%%%%%%%%%%%%%%%%%%%%%%%%%%%%%%%%%%%%%%%%%%%%%%%%%%%%%%%%%%%%%%%%%%

The starting point for this work was the observation
that there is a certain duality relationship between
the divide-and-conquer compression schemes of
Gupta-Verd\'{u}-Weissman (GVW) in \cite{gupta-et-al:pre:09},
and certain lossy Lempel-Ziv schemes based on a 
fixed-database as in \cite{kontoyiannis-lossy1-1}.
To explore this duality, LLZ, a new (non-universal)
lossy LZ algorithm was introduced, and it was shown
to be asymptotically rate-distortion optimal. 
To combine the low-complexity advantage of GVW
with the low-memory requirement of LLZ, a hybrid
algorithm, called HYB, was then proposed, and its properties
were explored both theoretically and empirically.

The main contribution of this short paper is the 
introduction of memory considerations in the 
usual compression-complexity trade-off.
Building on the success of the GVW algorithm,
it was shown that the HYB scheme
simultaneously achieves three goals: 
1. Its rate-distortion performance 
can be made arbitrarily close to the fundamental 
rate-distortion limit; 2. The encoding
complexity can be tuned in a rigorous manner
so as to balance the trade-off of encoding complexity 
vs.\ compression redundancy; and 3. The memory
required for the execution of the algorithm
is much smaller than that required by GVW,
a difference which may be made arbitrarily
large depending on the choice of parameters.

Moreover, empirically, for blocklengths of the order 
of thousands, the HYB scheme appears to outperform 
existing schemes
for the compression of simple memoryless sources
with respect to Hamming distortion.

Lastly, we briefly mention that the results presented 
in this paper can be 
extended in several directions.
First we note that the finite-alphabet assumption
was made exclusively for the sake of simplicity 
of exposition and to avoid cumbersome technicalities.
While keeping the structure of all three algorithms 
exactly the same, this assumption 
can easily be relaxed, at the price of longer,
more technical proofs, along the lines of 
arguments, e.g., in
\cite{yang-kieffer:2}%
\cite{kontoyiannis-lossy1-1}\cite{kontoyiannis-red:00}%
\cite{gupta-et-al:pre:09}.
For example, Theorem~4 of \cite{gupta-et-al:pre:09}
which gives precise performance and complexity bounds
for the GVW used with general source and reproduction
alphabets and with respect to an unbounded distortion
measure, can easily be generalized to HYB.
Similarly, Theorem~5 of \cite{gupta-et-al:pre:09} 
which describes the performance of a universal
version of GVW can also be generalized to the
corresponding statement for a universal version of HYB
(with obvious modifications), although, as noted
in \cite{gupta-et-al:pre:09}, the utility of that result
is purely of theoretical interest.

\newpage

\section*{Appendix}
%%%%%%%%%%%%%%%%%%%%%%%%%%%%%%%%%%%%%%%%%%%%%%%%%%%%%%%%%%%%%%%%%%

\subsection*{Proof of Theorem 1.}
%%%%%%%%%%%%%%%%%%%%%%%%%%%%%%%%%%%%%%%%%%%%%%%%%%%%%%%%%%%%%%%%%%
Recall that,
under the present assumptions,
the rate-distortion function
$R(D)$ is continuous, differentiable,
convex and nonincreasing \cite{berger:book}\cite{csiszar:book}. 
Given $D\in(0,\Dmax)$ and $\delta>0$, 
assume without loss of generality
that $D+\delta<\Dmax$;
then we can choose
$\gamma>0$ according to
$R(D+\delta)=R(D)-\gamma/2$,
so that 
$\Dbar=D+\delta$.
[As it does not change the asymptotic
analysis below, we take $\alpha>0$
fixed and arbitrary.]
Then the distortion part of the 
theorem is immediate 
by the construction of the
algorithm.

Before considering the rate, 
we record two useful asymptotic
results for the match-lengths $L_{\ell,k}$.
Let $R=R(D)+\gamma$, and 
$m=m(\ell)=\lfloor2^{\ell R}\rfloor+\ell-1$
as in (\ref{eq:m}).
Then \cite[Theorem~23]{dembo-kontoyiannis:wyner} 
immediately implies that,
$$\lim_{\ell\to\infty}\frac{L_{\ell,1}}{\log m(\ell)}
= \frac{1}{R(\Dbar)}\;\;\;\;\mbox{w.p.1.}$$
Moreover, for any $\epsilon>0$, 
the following more precise
asymptotic lower bound on $L_{\ell,1}$ holds: As
$\ell\to\infty$,
\be
(\log m(\ell))\Pr\left\{L_{\ell,1}\leq\frac{\log m(\ell)}{R(\Dbar)+\epsilon}
\;\Big| \;X_1^n \right\}
\;\rightarrow\; 0\;\;\;\;\mbox{w.p.1.}
\label{eq:tails}
\ee
The proof of (\ref{eq:tails}) is a straightforward
simplification of the proof of 
\cite[Corollary~3]{kontoyiannis-lossy1-1},
and therefore omitted.

Now let $\epsilon>0$ arbitrary.
The encoder parses the message $X_1^n$ into $\Pi_\ell$ 
distinct words $Z^{(k)}$, each of length $L_{\ell,k}$.
We let $N=(\log m(\ell))/(R(\Dbar)+\epsilon)$ and
following \cite{wyner-ziv:3} we assume,
without loss of generality, that
$N$ is an integer and that the last phrase 
is complete, i.e., 
$$Z^{(\Pi_\ell)}\;\mbox{ has length }\;L_{\ell,\Pi_\ell}.$$

To bound above the rate obtained by LLZ, we
consider phrases of different lengths separately.
We call a phrase $Z^{(k)}$ {\em long} if its length 
satisfies $L_{\ell,k} > N$,
and we call  
% and write $\Pi_\ell'$
% for the number of long phrases in the parsing of
% $x_1^n$. Otherwise 
$Z^{(k)}$ {\em short} otherwise.
% and the number of short phrases denoted by $\Pi_\ell''$. 
Recalling 
(\ref{eq:ell-bound}), the total description
length of the LLZ can be 
broken into two parts as,
\be
\Lambda(X_1^n)
&\leq&
	\sum_{k:\;Z^{(k)}\;\mbox{\footnotesize is short}}
	\Big[
	\lceil\log((1+\alpha)\ell)\rceil  
	+
	\lceil L_{\ell,k}\log |\Ahat|\rceil
	\Big]
	\nonumber\\
&\;& + \sum_{k:\;Z^{(k)}\;\mbox{\footnotesize is long}}
	\Big[
	\lceil\log((1+\alpha)\ell)\rceil  
	+ \lceil\log m\rceil
	\Big].
	\label{eq:total-bound}
\ee
For the first sum we note that,
by the choice of $m(\ell)$ and the
definition of a short phrase,
each summand is bounded above 
by a constant times $N$, at least
for all $\ell$ large enough; therefore,
the conditional expectation 
of the whole sum 
given $X_1^n$
is bounded by,
\ben
E\left\{ C_1\,N\,
	\sum_{k=1}^{\Pi_\ell}
	{\mathbb I}_{\{L_{m,k}\leq N\}}
	\,\Big|\,X_1^n\right\}
\leq
C_2 \log m(\ell) \, n\, \Pr\left\{L_{m,1}\leq
	\frac{\log m(\ell)}{R(\Dbar)+\epsilon}
	\;\Big|\;X_1^n\right\},
\een
where 
${\mathbb I}_F$ denotes
the indicator function of an event $F$,
and the inequality follows by
considering not just all $k$'s, but all the possible positions
in $X_1^n$ where a short match can occur. Dividing by $n$
and letting $n\to\infty$, from (\ref{eq:tails}) we get that
this expression converges to zero w.p. 1, so that 
the conditional expectation of the first term 
in (\ref{eq:total-bound}) also converges to zero, 
w.p.1.

For the second and dominant
term in (\ref{eq:total-bound}),
let $\Pi'_\ell$ be the number of long phrases $Z^{(k)}$. 
Since each long $Z^{(k)}$ 
has length $L_{m,k}\geq N$, we must have 
$N\Pi'_\ell \leq n$, so that 
\be
\frac{\Pi'_\ell}{n} \leq \frac{R(\Dbar)+\epsilon}{\log m(\ell)}.
\label{eq:bad}
\ee
Also, by the definition of $m(\ell)$,
for all $\ell$ large enough (independently
of $n$), we have,
$$\log((1+\alpha)\ell)\leq \epsilon\log m(\ell).$$
Therefore, the second sum in 
(\ref{eq:total-bound}) 
can be bounded above by,
$$
	\Pi'_\ell\,(1+\epsilon)\log m(\ell)
\leq
	n(1+\epsilon)(R(\Dbar)+\epsilon).
$$
Combining this with the fact that the first term in 
(\ref{eq:total-bound}) vanishes, immediately yields
\ben
\limsup_{\ell\to\infty}\limsup_{n\to\infty}\;
E\left\{\frac{\Lambda(X_1^n,D,\ell)}{n}\;\Big|\,X_1^n\right\}
\;\leq\; (R(\Dbar)+\epsilon)(1+\epsilon)
\;\;\;\;\mbox{w.p.1},
\een
and since $\epsilon>0$ was arbitrary we get the first
claim in the theorem.
Finally, the second claim follows from the
first and Fatou's lemma. 
\qed

\subsection*{Proof of Theorem 2.}
%%%%%%%%%%%%%%%%%%%%%%%%%%%%%%%%%%%%%%%%%%%%%%%%%%%%%%%%%%%%%%%%%%

The proof of the theorem is based on Lemma~1 below,
which plays the same role as 
\cite[Lemma~1]{gupta-et-al:pre:09}
in the proof 
of \cite[Theorem~1]{gupta-et-al:pre:09}. The rest of 
of the proof is identical, except for the fact that 
we do not need to invoke the law of large numbers, 
since here we do not claim that 
the probability of excess distortion goes 
to zero.
\qed

Before stating the lemma, we define the
following auxiliary quantities: $D_1=D/2$,
$K(D)=(D-D_1)/(R(D_1)-R(D))$, 
$$C(D)=\min\Big\{
\frac{K(D)^2}{8\Dmax^2},\frac{1}{32(R'(D/2)\Dmax)^2},\frac{1}{4}
\Big\},$$
and,
$$\hat{\epsilon}=
\min\Big\{\frac{\exp\{16C(D)\}}{3(\Dmax-D)},
3e^{-1}(\Dmax-D)
\Big\}.$$

\medskip

\noindent
{\bf Lemma 1. } 
Consider a memoryless source $\{X_n\}$
to be compressed at target distortion
level $D\in(0,\Dmax)$. Then
for any $0<\epsilon<\hat{\epsilon}$,
the HYB algorithm with parameters
$0<\gamma<\hat{\gamma}$ and
\be
\ell=\left\lceil
\frac{1}{C(D)\gamma^2}\log\frac{3(\Dmax-D)}{\epsilon}
\right\rceil,
\label{eq:ellDefn1}
\ee
when applied to a single block
$X_1^\ell$
achieves rate $R=R(D)+\gamma$, 
and its expected distortion is less 
than $D+\epsilon$.

\medskip

\noindent 
{\em Proof. }
Given $\epsilon>0$, choose a positive $\epsilon'<\epsilon$
such that,
$$
\frac{\epsilon'}{C(D)}\log\frac{3(\Dmax-D)}{\epsilon'}<\epsilon.$$
Now follow the proof of 
\cite[Lemma~1]{gupta-et-al:pre:09}
with $\epsilon'$ in place of $\epsilon$,
until the beginning of the computation
of the probability of excess distortion.
The key observation is that, for HYB,
this probability can be bounded above by
the excess-distortion probability with
respect to a random codebook with 
$$
\frac{1}{\ell}2^{\ell R(\Dbar)}=
2^{\ell (R(D)+\gamma-\frac{\log \ell}{\ell})}
$$
words, by just considering possible
matches starting at positions
$i=1,\ell+1,2\ell+1,\ldots$,
making the corresponding potentially
matching blocks in the database independent.
Therefore, following the same computation,
the required probability
can be bounded above as before by,
\be
2(2^{-\ell C(D)\gamma^2})+\ell 2^{-\ell\gamma/4}.
\label{eq:prob}
\ee
The first term is bounded above by,
$$\frac{2\epsilon'}{(\Dmax-D)},$$ 
as before,
and in order to show that
the expected distortion is less than
$\epsilon$ it suffices to show that 
the last term satisfies,
\be
(\Dmax-D)\ell2^{\ell (R(D)+\gamma)}<\epsilon/3.
\label{eq:target}
\ee
Substituting the
choice of $\ell$ from (\ref{eq:ellDefn1}),
it becomes,
$$\frac{(\Dmax-D)}{C(D)\gamma^2}\log\Big(\frac{3(\Dmax-D)}{\epsilon'}\Big)
2^{-\frac{1}{4\gamma C(D)}\log(3(D_{\rm max}-D)/\epsilon')},$$
and since $\gamma$ is restricted to be less than one,
this can in turn be bounded above, 
uniformly in $\gamma\in(0,1)$,
by its value at $\gamma=1$. [To see that, note that
the function $f(x)=Ax^2\exp\{-Bx\}$ is increasing for
$x<2/B$ and decreasing for $x>2/B$. By our choice
of $\hat{\epsilon}$, the maximum above
is achieved at the point $x=1/\gamma=1$.] Therefore,
noting also that $4C(D)\leq 1$, 
this term is bounded above by,
$$\frac{(\Dmax-D)}{C(D)}\log\Big(\frac{3(\Dmax-D)}{\epsilon'}\Big)
2^{-\log(3(D_{\rm max}-D)/\epsilon')},$$
which, after some algebra, simplifies to,
$$\frac{\epsilon'}{3C(D)}\log\Big(\frac{3(\Dmax-D)}{\epsilon'}\Big),$$
and this is less than $\epsilon/3$ by
the choice of $\epsilon'$.
This establishes (\ref{eq:target}) and 
completes the proof of the lemma.
\qed

\subsection*{Proof of Theorem 3.}
%%%%%%%%%%%%%%%%%%%%%%%%%%%%%%%%%%%%%%%%%%%%%%%%%%%%%%%%%%%%%%%%%%
Taking $c>0$ arbitrary, we let, as in the proof
of \cite[Theorem 3]{gupta-et-al:pre:09},
$$\ell(n)=\left\lceil\frac{\log g(n)}{R(D)+c}\right\rceil
\;\;\;\;\mbox{and}\;\;\;\;
\gamma(n)=\sqrt{
\frac{\log \ell(n)}{\ell(n)}
}.
$$
For each $n$ we use HYB with the corresponding parameters;
the rate result follows from the construction of the
algorithm, which, at blocklength $n$, 
has rate no larger than,
$$R(D)+\gamma(n)\to R(D)\;\;\;\;\mbox{bits/symbol},$$ 
as $n\to\infty$.

Regarding the distortion, 
equation (\ref{eq:prob}) in the proof
of Theorem~2 shows that that
the probability of the event that 
the distortion of the $i$th block 
will exceed $D$ is bounded above by,
$$2(2^{-\ell(n) C(D)\gamma(n)^2})+\ell(n) 2^{-\ell(n)\gamma(n)/4}.$$
It is easily seen that,
for large $n$, this is dominated
by the second term, 
$$\ell(n) 2^{-(1/4)\sqrt{\ell(n)\log\ell(n)}}.$$
Therefore, the distortion of any one
$\ell$-block is bounded above by,
$$D+
\Dmax\ell(n) 2^{-(1/4)\sqrt{\ell(n)\log\ell(n)}}.$$
Noting that the excess term goes 
to zero as $n\to\infty$, it will still
go to zero when averaged out over all
$n/\ell(n)$ sub-blocks, and, therefore,
the expected distortion over the whole
message $X_1^n$ will converge to $D$.

Finally, the complexity results are 
straightforward by construction;
see the discussion in \cite[Section~II-A]{gupta-et-al:pre:09}.
\qed

% \newpage

\section*{Acknowledgments}
%%%%%%%%%%%%%%%%%%%%%%%%%%%%%%%%%%%%%%%%%%%%%%%%%%%%%%%%%%%%%%%%%%

We thank Sergio Verd\'{u} for sharing with us preprints
of \cite{gupta-verdu:pre} and \cite{gupta-et-al:pre:09}.

% \bibliography{../../latex/ik}

\end{document}